# Improvement of WAVEWATCH III output through a wind speed modification based in boundary layer temperature variability.

S Gremes Cordero, E. Rogers, Y. Fan

**Abstract**

We present here an empirical method aimed at decreasing the error in the significant wave height ($H_s$) calculated through the WaveWatch III® model (WW3). The errors are calculated as the difference between the modeled and the in-situ observed measurement. We hypothesize that this error would be reduced if the model used a well-calibrated method to account for thermal variations within the surface boundary layer (SBL).

We compared then the $H_s$ error for 2015 to the air-sea temperature difference (ASTD) in order to find a potential relationship among them. The statistical analysis performed show a clear correlation between these variables, hinting for the need of introducing a correction for SBL stability based on a linear relationship between the error and the air-ocean temperature difference. We developed a wind speed adjustment, which allows ASTD effects to be introduced into the model forcing. Such adjustment proved effective in reducing the bias in $H_s$ for negative values of ASTD, though with limited success, suggesting the need for extending the approach, e.g. by using an iterative cycle of subsequent corrections.

## 1 – Introduction

The prediction capability of environmental models depends largely on the reliability of the algorithms and parameterizations used to represent the physical forcing involved, and their multifaceted interactions. In particular, coupled models of global circulation rely on the correct representation of the properties' exchange at the ocean surface.

It is natural then that much attention is directed towards the parameterization of the microscale processes occurring within and around the oceanic surface boundary layer (SBL, from now on). In particular, the interaction between waves, currents and turbulence are among the most important processes, and the most complex, to understand and correctly formulate. One of the most challenging problems is to solve the wind profile within the SBL, influenced by the local temperature variability and atmospheric stability.

These interconnected effects had been gradually implemented into different versions of the WAVEWATCH III® model (WW3). The WW3 is a global model capable of forecasting several surface waves characteristics and has been proven its great capability to predict wave height and other wave properties successfully for several years. Since its creation, the WW3 has been subject to many improvements, resulting in a series of 'switches' that can be enabled according to the user needs (WAVEWATCH III Development Group 2016, hereafter W3DG). Indeed, the sources (and sinks) of energy had been partitioned according to their effect in the system allowing for several alternatives (Zieger, 2015; Liu et al, 2019; Rogers and Orzech, 2013; Tolman and W3DG, 2014). In particular, the switches STAB (0-4) handle the introduction of stability corrections into the model (see section 3).

Changes in air density, related to temperature variability and wind gustiness are also influential on the stability of the SBL (Abdalla, 1991; Abdalla and Cavaleri, 2002; Pietersen and Hams, 2013; Young, 1998). These effects were introduced into the WW3 as described in Abdalla and Bilot (2002). The parameterization of these effects into the WW3 model and the corresponding sources (switches) are explained in detail in Ardhuin et al. (2010) and involve considering the input term dominated by the exponential wind-wave growth (*Sin*), plus a wave-ocean interaction term that contains the dissipation (*Sds*).

Tolman (2002) addressed the impact of the air and sea temperature difference within SBL on the wave growth and its parameterization into the WW3. He proposed the introduction of a stability parameter, S:

$$S = \frac{h\,g}{u_h^2} \frac{T_a - T_s}{T_o} \qquad (1)$$

with *u* and *Ta* as wind speed and air temperature at height *h* and *Ts* and *To* as the sea and reference temperature, respectively. His study suggest that the model error was a weak function of the atmospheric stability and pointed out how the apparent underestimation of wave heights in unstable conditions, consistent with the observed higher growth rates in such conditions.

In this context we decided to investigate the relationship between the significant wave height error (obtained as the difference between the model and the in-situ observations), and the air-sea temperature difference. In order to reduce the bias or error in $H_s$, we propose to use a method similar to Tolman's, but based solely in the air-sea thermal difference (ASTD). The plan is to adjust the wind speed to account for the effect of this variability in the SBL. The main advantage of our approach over Tolman's considerations, resides in avoiding the 'reference temperature', which make the approximation ambiguous and not globally applicable. We are focused in the temperature profile within the SBL, and the effects of its variability over the boundary layer dynamics and especially over the wind. These effects directly relate to the stability of the SBL and hence to all ocean surface dynamics and processes related, like surface wave growth, gas exchange, mass and momentum fluxes, etc. We particularly concentrate in the variations of ASTD in relation to the wave growth forecasts.

Therefore, we compared the output from the WW3 with in-situ data obtained through the National Data Buoy Center (NDBC) platforms. Three data sets of modeled $H_s$ were utilized for this research: the 'original' data (case 1) with winds from an atmospheric model; the data corrected with the stability correction integrated within the WW3 script (case 2); and a data set resulting from modifying the wind according to our adjustment (case 3). In the first case, a stability correction is not allowed, while in the other two cases a suitable switch allowing for stability correction based in air density variability is used (see description in section 3).

In the following, section 2 presents a review of the surface boundary dynamics and its governing equations with emphasis in our focus; section 3 introduce the details of the data sets analyzed, and section 4 the methods used for the analysis. The results and discussion appear in section 5, followed by applications and finally the conclusions.

**2 – The Surface Boundary layer and the neutral equivalent wind speed**

We can define the atmospheric surface layer or oceanic surface layer, as the bottom 10% of the atmospheric boundary layer. Within this surface boundary layer (SBL from now on), far from the first mm viscous layer, momentum and mass fluxes can be defined as:

$$\vec{\tau} = \rho \left( -\overline{w'u'}\,\hat{\imath} - \overline{w'v'}\,\hat{\jmath} \right) \quad \text{and} \quad F_x = \rho\,\overline{w'x'} \qquad (2)$$

respectively. Primes indicates turbulent fluctuations and the over bars represent a suitable average in time (Drennan, 2006). The wind velocity components are represented by $u', v', w'$ (horizontal in- line with the mean wind, horizontal cross-wind and vertical, respectively), $\rho$ is the air density, and $x'$ is the mixing ratio of any substance with respect to dry air.

Most of the analysis done over the turbulent sublayer is based on the theories of Obukhov (1946) and Monin-Obukhov (1954) about flow similarity. Assuming stationary and homogeneous conditions within the surface layer, the vertical momentum flux is considered constant and proportional to the square of the so-called friction velocity. The friction velocity is introduced as a velocity scale in the Monin-Obukhov (MO) theory (Monin and Obukhov, 1954; Obukhov, 1946). Assuming that turbulence in the surface layer (far from the very near-surface region, where molecular processes are also important) is generated by only surface shear and buoyancy, the MO theory determine that the gradients and scaling parameters are related by means of universal dimensionless gradient functions as:

$$\frac{\partial U}{\partial z} = \frac{u_*}{\kappa z} \Phi_u\left(\frac{z}{L}\right) \quad \text{and} \quad \frac{\partial X}{\partial z} = \frac{x_*}{\kappa z} \Phi_x\left(\frac{z}{L}\right) \quad (3)$$

for momentum and mass respectively, where $z$ is the mean height above the surface and $L$ is the Obukhov scale height, $L = -u_*^3 / \kappa g (\overline{w'\theta'} / T_0 + 0.61\,\overline{w'q'})$, with $\kappa = 0.4$ as the Von Kármán constant, $g$ the gravitational constant, $q'$ is the humidity fluctuation (equivalent to $x'$ above), $\theta'$ the potential temperature fluctuation and $T_0$ a reference absolute temperature (Drennan, 2006; Csanady, 2001). The Obukhov scale height represents the effect of density stratification within the SBL. The ratio between the measurements' height above the ocean and the Monin-Obukhov length is called the stability parameter ($z/L = \zeta$) and represents a balance between buoyant and mechanical effects (Large and Pond, 1982; Monin and Yaglon, 1971; Businger et al., 1971)

Integrating (3) from the surface to some height z in the constant flux layer, the velocity profile can be obtained in terms of the gradient functions $\Phi$. Deardoff (1968) proposed, for the stable case ($L > 0$),

$$\Phi(z/L) = 1 + \beta\,\zeta \quad (4)$$

And for the unstable case ($L < 0$),

$$\Phi(z/L) = 1/(1 - \alpha\,\zeta)^{1/4} \quad (5)$$

where $\alpha, \beta$ are determined empirically.

Louis (1979) utilized Businger's functions to solve the Obhukov height analytically, and the gradient functions, obtaining:

$$\Phi(\zeta) = \ln\left[\left(\frac{1+x}{2}\right)^2 \left(\frac{1+x^2}{2}\right)\right] - 2\arctan x + \frac{\pi}{2} \tag{6}$$

with

$$x = (1 - \gamma_m \zeta)^{1/4} \quad \text{for unstable conditions and}$$

$$\Phi(\zeta) = -\beta\zeta \quad \text{for stable conditions.}$$

An application for simple cases appears in Louis et al. (1982).

In neutral conditions, i.e., with no vertical density gradient, the buoyancy effects can be disregarded (Young, 1998; Donelan, 1990), and one can obtain the well-known logarithmic law for the wind velocity profile:

$$\frac{U_z}{u_*} = \frac{1}{\kappa} \ln \frac{z}{z_0} \tag{7}$$

It is customary to eliminate the dependence on stability and height by considering the neutral drag coefficient at a reference height of 10 m. The same consideration is applied to the humidity fluxes as well as to the wind speed (and it is called $U_{10N}$).

Note that assuming MO theory, the terms in (3) are expected to be universal functions of $\zeta$. With little experimental data available on either horizontal or vertical variations of the gradient functions, they are usually combined as a single transport (or imbalance) term. This term either is then assumed null (Large and Pond, 1981), or is determined empirically (Dupuis et al., 1997). When $\Phi_u = 1$, that is, for neutral conditions, we can obtain the well know 'Law of the wall' definition for the dissipation of energy, $\varepsilon = u_*/\kappa z$.

In the experimental field, Young (1998, 2018) calculated the role of the atmospheric stability in wind wave growth by studying the changes in the planetary boundary layer under neutral, stable and unstable conditions. He used for his purpose in situ observations of air and seawater temperature and wave height collected on Lake George, with shallow uniform depth and over steady wind conditions. He showed how the wind wave growth is enhanced in unstable conditions (when water is warmer than the air) and diminished during stable conditions (water colder than air). In neutral conditions, when there is no vertical density gradient within the surface boundary layer (SBL), the airflow can be described by a well-known logarithmic function, as described previously (Young, 1998, Donelan, 2008).

## 3 – The data

### 3.1- In-situ observations

The National Data Buoy Center provides real time and historical data from ocean meteorological buoys, as wind speed and direction, atmospheric pressure, air temperature (Ta), dew point, etc. The data available include measurements obtained from C-MAN and moored buoys, with different mast height, and platform shape and size (for more details check https://www.ndbc.noaa.gov ). The data is obtained every 10 minutes and transmitted to the NDBC for quality control and storage.

Of the hundreds deployed around the USA we used 56 buoys, located at least 50 km offshore (Figure 1). Here we use measurements obtained during the whole year of 2015. The model values are linearly interpolated to the observations.

### 3.2 - WAVEWATCH III ®model output

The WAVEWATCH III ® is a wind-wave modelling framework managed by the National Center for Environmental Prediction (NCEP). The model development relies on several institutions, including the NCEP and the Naval Research Laboratory. Predecessor codes were developed at the Delft University of Technology and the NASA Goddard Space Flight Center. For full system details, the source code is fully documented in Tolman, (2010 and 2014) and Liu et al. (2019) and previous publications (Tolman, 1991, 1992, 1997, 2002, 2010, Tolman and Chalikov, 1996; Tolman and Grumbine, 2013).

The current model version (5.16), present some differences with previous versions mostly related to run time optimization, sea ice mask improvements, options for nonlinear wave-wave interaction source terms, and additional sea-state dependent stress-calculations (W3DG, 2016). Up to date information on this model can be found on the WAVEWATCH III web page, http://polar.ncep.noaa.gov/waves/wavewatch.

The output of WAVEWATCH III consists of the traditional frequency-direction spectrum and different spectra can be derived by means of straightforward Jacobian transformations. However, within WAVEWATCH III the wavenumber-direction spectrum has been selected because of its invariance characteristics with respect to physics of wave growth and decay for mutable water depths.

For the input sources, a term 'S' is used to represent the net effect of sources and sinks for the spectrum. They are divided in groups to allow the user the addition of different effects, as well as their parameterization, as flux calculation type, non-linear interactions, bottom friction, etc. When no term for input and dissipation is used, the switch or package is 'STAB0'; when the stability correction is enabled is called STAB3 and considers the input term dominated by the exponential wind-wave growth ($Sin$), plus a wave-ocean interaction term that contains the dissipation ($Sds$).

The wind speed forcing the WW3 is a product obtained from the NAVGEM (U.S. Navy Global Environmental Model), a high-resolution weather prediction model, which provides synoptic forcing at the ocean surface in 3hs-intervals. Indeed, NAVGEM provided the input needed for the WW3 runs: wind speed at 10 m, sea surface temperature, and air temperature with nearly a quarter-degree resolution. The wind speed includes a stability correction applied for the difference in height between the first model grid level and the 10 m height used here. This correction is based in the Monin-Obukhov similarity theory, as explained in section 2, and includes a simple parameterization of the gradient functions, following Louis et al. (1982).

The WW3 output used for our statistical analysis are global hourly values of $H_s$ at quarter-degree spatial resolution. We interpolated linearly to obtain matched-up observations from the NDBC buoys. We studied three cases:

Case 1: No additional stability correction within SBL (STAB0);

Case 2: Stability correction added through STAB3 switch, input ASTD from NAVGEM;

Case 3: Allowed a stability correction as in case 2), with winds adjusted according to the ASTD-$H_s$ curve (section 5) before the run.

4 – Methods

The buoy data, obtained every 10 min, was resampled to match the temporal resolution of the model thus obtaining hourly data. Spatial interpolation was used to obtain model data at the given buoy locations (matchups). The time series for each match-up extended for the whole year with few exceptions (3 buoys out of 56).

We calculated the model 'error' or 'bias' in $H_s$, by subtracting the in-situ observation from the $H_s$ calculated through the model, for all cases. Furthermore, we calculated the ASTD from the NAVGEM model at the same locations, for the year considered here.

The statistical analyzes performed here consisted mainly in two types of calculations:

- Density scatter plots of ASTD (in °C) vs $H_s$ error (in m) with a colorbar representing the density of observations (match-ups) throughout the year;

- Histograms of 'binned' ASTD values (1°C) vs $H_s$ error (m). The colors in the figure denote each different bin histogram. The ratio between the number of observations in the bin and the max value of observations appear in the left vertical axis.

For our wind adjustment (section 5), we fit the observations (which are mean values of $H_s$ per bin of ASTD) to a polynomial function, allowing us to extend the range of possible ASTD and $H_s$ values.

## 5 – Results and discussion

### *5.1 – Correlations - Case 1*

The scatter distribution of the data in Figure 2 shows large scatter for most of the bins, with a center (in red) shifted from zero, and negative $H_s$ error corresponding the negative values of ASTD. In addition, the densest points appear at a positive value of $H_s$ error, and ASTD.

The histograms in Figure 3, reveal a clear correlation between the $H_s$ error and the ASTD, with peaks at -.2 and +.1. Ideally, the maximum error for each bin would be zero.

### *5.2 – Introducing ASTD effects - Case 2*

Given the clear correlation displayed between $H_s$ error and the ASTD, we investigated the introduction of thermal stability into the model using an existing method (STAB3) and also through a new empirical method proposed here.

The existing method of including this effect into the calculations is to input the ASTD values obtained from the NAVGEM model into the WW3 model, with the STAB3 switch turned on (allowing for thermal stability corrections) as described in Abdalla and Bidlot (2002, 2010).

The $H_s$ error calculated as modeled-observed value was then compared to ASTD and is presented in Figure 4a. We find more scattered values than in the previous case, with a maximum density not located at zero error as in the previous case. In addition, the histograms presented in Figure 4b do not indicate any improvement for either negative or positive values of ASTD, and the negative values lost the correlation with $H_s$.

### *5.3– Implementing error reduction (case 3)*

We explored then a different method of adding the thermal difference effects, by designing an adjustment based in the empirical observations. First, the mean $H_s$ error was calculated for each bin, as appear in Table 1, together with the quantity of occurrences for each bin of ASTD. The plot of $H_s$ versus ASTD is represented in Fig. 5. A polynomial fit for these observations was not a straightforward solution, as most of the data was concentrated between bin -10 and +1 C. We then decided to use a step function for any value compressed between those 2 limits, a fifth order function for the data with ASTD smaller than -9, and a second order fit was used to represent the curve fit. This curve is solely a representation of this data set and can be used as a reference to know the value of $H_s$ for any given value of ASTD, that is, to expand the results obtained for the NDBC locations to the rest of the ocean.

An indirect way of introducing the thermal stability effects would be to adjust the wind speed in such a way that the thermal difference between the top and the bottom of the SBL is allowed to affect the wind profile, and hence, the derived wind speed at 10 m. The procedure that we used is as follows.

In a first approximation, we have a mean value of $H_s$ error for each degree of the ASTD, or bin. To obtain a relationship between the $H_s$ error and the wind, we use the Pierson-Moskowitz general description of energy distribution, which states:

$$H_s = C * U_{10}^2 \qquad (5)$$

where $C = \beta/g = 0.024598$ and $\beta$ = 2.6812 $\sqrt{0.0081}$. The limitations of this approach will be described in the next section.

We can estimate the variations in $H_s$ due to wind speed fluctuations as the ratio $\Delta H_s/\Delta U$, in which case following (5),

$$\Delta H_s/\Delta U = 2 * C * U_{10} \qquad (6)$$

and
$$\Delta U = \Delta H_s/(2 * C * U_{10}) \qquad (7)$$

In this way, our 'adjusted' wind speed becomes:

$$U_{adj} = U_{10} - \Delta U = U_{10} - \Delta H_s/(2 * C * U_{10}) \qquad (8)$$

We can apply this relationship to each wind component by considering:

$$\frac{U_{adj}}{U_{10}} = \frac{u_{x,adj}}{u_x} = \frac{u_{y,adj}}{u_y} \tag{9}$$

Once the wind speed is adjusted in this way, we run the model with no stability correction as in the first case, just with the corrected winds. We calculated the $H_s$ error in this case again as modeled minus observation for each match up available and obtained the scatter plot in Figure 6a. We observe that even if some scatter is still apparent, the maximum density of observations is now clearly centered on zero error and the vertical spread is reduced (see yellow and green shades toward the center of the distribution).

The histograms in this case (Figure 6b) displayed a reduced pick, i.e., there are less observations with errors in both positive and negative ASTD cases. Also, the range for the maximum and minimum values of the distribution (horizontal) is slightly reduced and shifted toward zero when compared with the two previous cases

The biases at each buoy site has been effectively reduced with our method as we can observe in Figure 7.

We need to acknowledge the implicit limitations involved in the use of the Pierson-Moskowitz (PM) theory. Indeed, the PM distribution of energy is very general, applicable in global large scales, and does not include smaller scale effects as for example the effects of swell, wind gustiness, air density, etc. While it is a limiting factor, our method reduced the bias more effectively than when considering winds gustiness and air density effects within the WW3.

Moreover, it is expected that the iterative cycle of recalculating error after each wind adjustment will result in an even reduced error in $H_s$.

6 – Alternatives and implications for the Earth System Prediction Capability (ESPC) system

A parallel experiment was performed by Jean Bidlot, from the ECMFW group, using the ERA5 forcing (the latest ECMWF reanalysis). The new configuration includes gustiness effects and air density variability and is forced by 10 m neutral winds. To understand the effects of his new implementation, he compared three outputs, 1- with regular winds, without gustiness and air density effects; 2- with neutral winds, without gustiness and air density effects; 3- with neutral winds, including gustiness and air density effects.

The model data has a temporal resolution of 6 hours. He compared the results initially to the same NDBC buoys we used, and to all the in-situ data they have available (about 5 times the previous quantity of observations).

His scatter plots with regular winds are quite identical to our findings. Most number of colocations are centered around -1C, indicating like in our tests, that most of the observations are obtained in unstable conditions (with negative ASTD). More interestingly, using neutral equivalent winds really seem to remove most of the air-sea temperature effect, noticeable reducing the $H_s$ bias. Adding gustiness and air density variability had a much more muted impact (Bidlot, personal communication).

The use of neutral equivalent winds implies that the effects of atmospheric stability are incorporated into the dynamics of the model, and the forcing wind has been already corrected for these effects. Would it be the case also for our ESPC system?

The ESPC system is a prediction model which outputs oceanic and atmospheric variables by coupling the NAVGEM model with the HYCOM model (HYbrid Coordinate Ocean Model). The WW3 is incorporated at the interface. The use of neutral equivalent winds to force the WW3 would enhance the wind speed in unstable conditions and reduce the wind speed in stable conditions (Table 1), as explained in section 2. It is then expected that the underestimation of $H_s$ in unstable conditions would be less remarkable, and the bias with observations would then be reduced. The contrary applies for stable conditions, which are less common in open ocean (Young et al, 2017)

To test this hypothesis, the output from WW3 within ESPC with neutral equivalent winds would be analyzed in the future.

## 7 – Summary and conclusions

We detected a clear correlation between $H_s$ error and ASTD when comparing the output from WW3 and the NDBC data for 2015 and analyzed three different cases.

For the first case analyzed (case 1), with no additional corrections to winds or boundary layer stability, the mean density of number of observations was not centered at zero error, and remarkable scatter appear in $H_s$ values when compared to ASTD. Furthermore, the $H_s$ histograms for each degree Celsius of ASTD showed the clear correlation with the $H_s$ error with maximum occurrences shifted from zero.

We proposed an indirect method of introducing thermal stability into the SBL by adjusting wind speed as to account for the ASTD variations. We used the observations to get the error $H_s$ values for each ASTD, and the PM relationship to translate the variations of $H_s$ into variations of wind. We compared the results from the original run (case 1), the results with the stability correction set within the model (STAB3), which allows for wind gustiness and air density variability (case 2), and the results obtained after our own adjustment to the wind speed (case 3). Our adjustment demonstrated to be the most successful in reducing biases. However, a more aggressive correction is needed, which we are planning to implement by an iterative process.

In addition, a comparison with the results of using neutral equivalent winds within the ESPC is necessary to assess alternative corrections.

**Figure captions**:

Figure 1: NDBC buoy positions. The colors correspond to the $H_s$ error (in m).

Figure 2: Scatter plot of $H_s$ error (in m) versus ASTD (in °C) for the original model output (case 1). Colors indicate the quantity of occurrences per 1 °C-bin.

Figure 3: Histograms of $H_s$ per bin of temperature (1 °C) for the same data set of Fig 2. The colors show the different bins. The y-axis indicates number of occurrences normalized by the number of observations per bin.

Figure 4: a) Scatter plot of $H_s$ error (in m) versus ASTD (in °C) for the case 2. Colors indicate the normalized occurrences per bin of 1 °C. b) Histograms of $H_s$ per bin of temperature for the same data set. The colors indicate the different ASTD bins. The y-axis indicates number of occurrences normalized by the number of observations per bin.

Figure 5. $H_s$ (in m) vs ASTD (in °C) observed (*), and polynomial fits (continuous line).

Figure 6. a) Scatter plot of $H_s$ error (in m) versus ASTD (in °C) for the case 3. Colors indicate the normalized occurrences per bin of 1 °C. b) Histograms of $H_s$ per bin of temperature for the same data set. The colors show the different ASTD bins. The y-axis indicates number of occurrences normalized by the number of observations per bin.

Figure 7. Biases in $H_s$ (m) for the matchups, for case 2 (a), and case 3 (b).

**Figures**

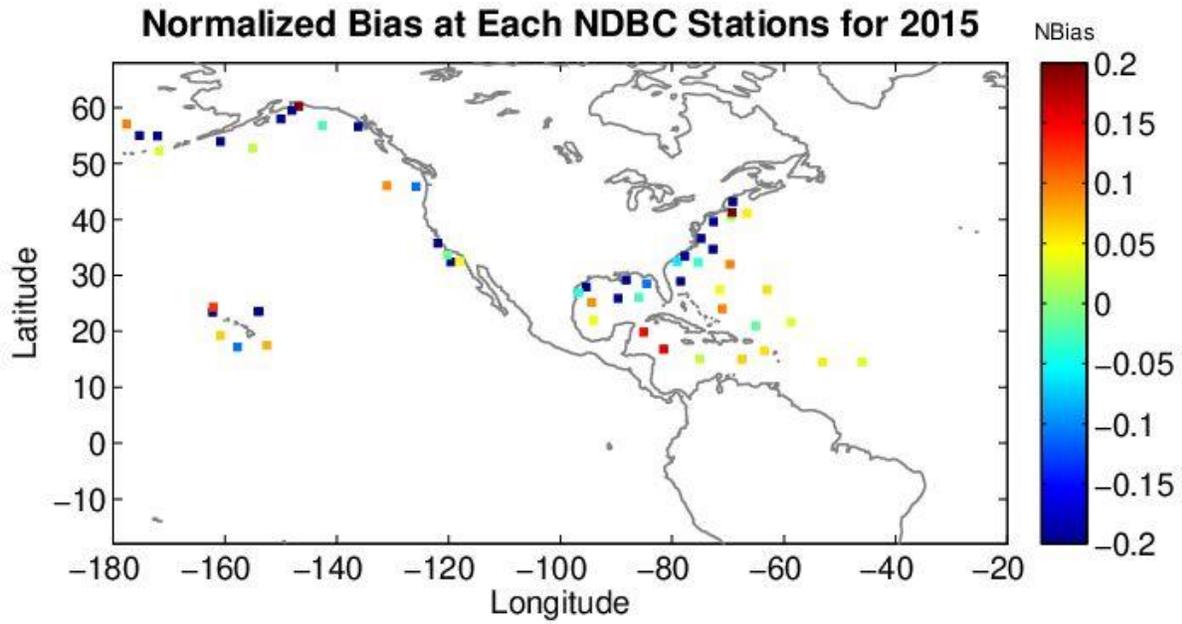

Figure 1: Normalized bias at the NDBC buoys for 2015. The colors correspond to the $H_s$ error or bias (in m).

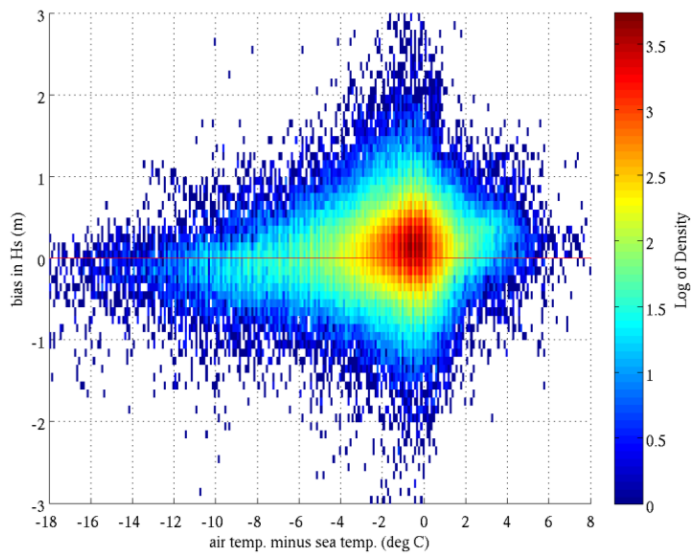

Figure 2: Scatter plot of $H_s$ error (in m) versus ASTD (in °C) for the original model output (case 1). Colors indicate the quantity of occurrences per bin of 1 °C.

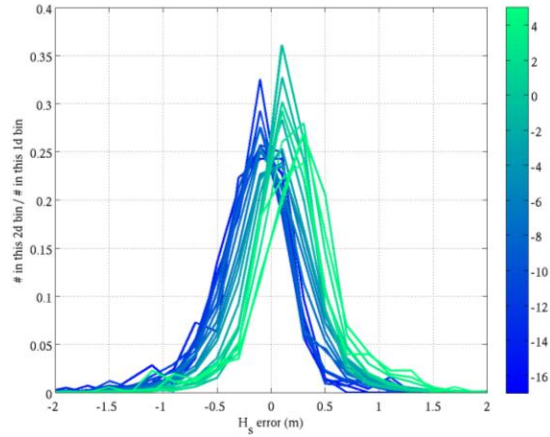

Figure 3: Histograms of $H_s$ per bin of temperature (1 °C) for the same data set of Fig 2. The colors show the different bins. The y-axis indicates number of occurrences normalized by the number of observations per bin.

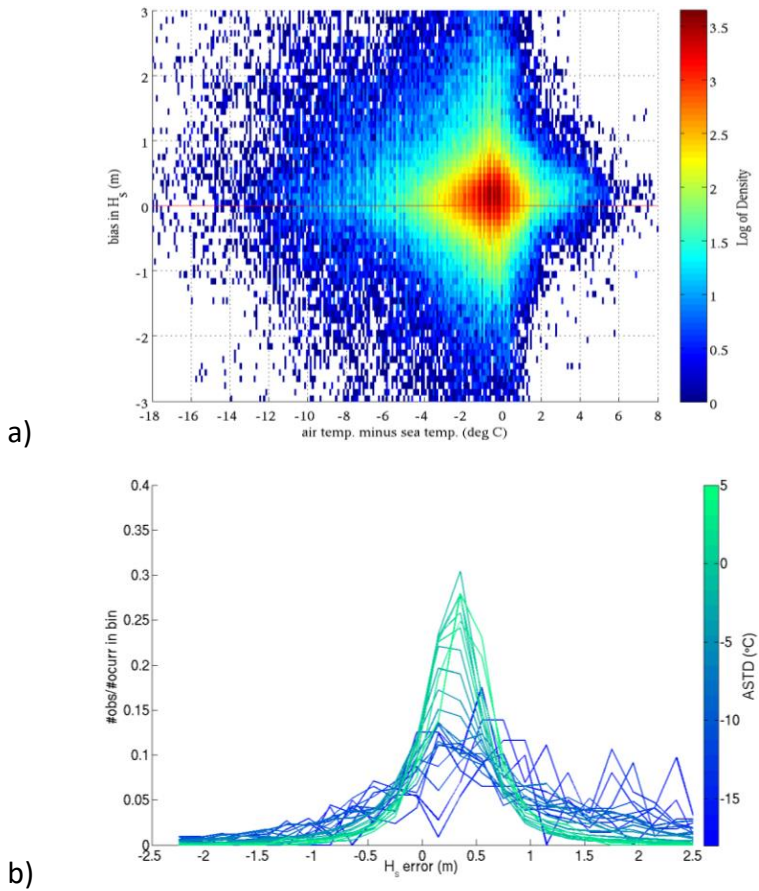

a)

b)

Figure 4: a) Scatter plot of $H_s$ error (in m) versus ASTD (in °C) for the case 2. Colors indicate the normalized occurrences per bin of 1 °C. b) Histograms of $H_s$ per bin of temperature for the same data set. The colors show the different temperature bins. The y-axis indicates number of occurrences normalized by the number of observations per bin.

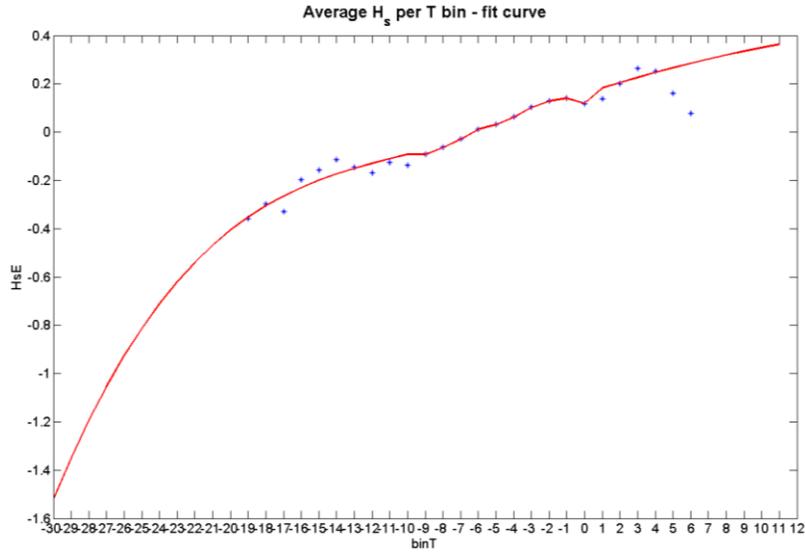

Figure 5. $H_s$ (in m) vs ASTD (in °C) observed (*), and polynomial fits (continuous line).

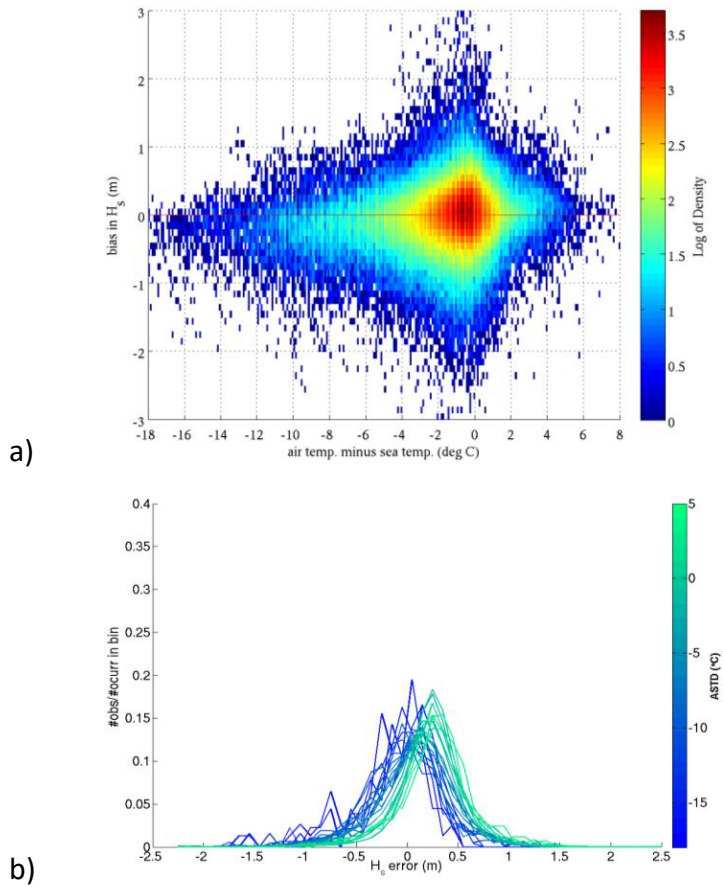

a)

b)

Figure 6. a) Scatter plot of $H_s$ error (in m) versus ASTD (in °C) for the case 3. Colors indicate the normalized occurrences per bin of 1 °C. b) Histograms of $H_s$ per bin of temperature for the same

data set. The colors show the different bins. The y-axis indicates number of occurrences normalized by the number of observations per bin.

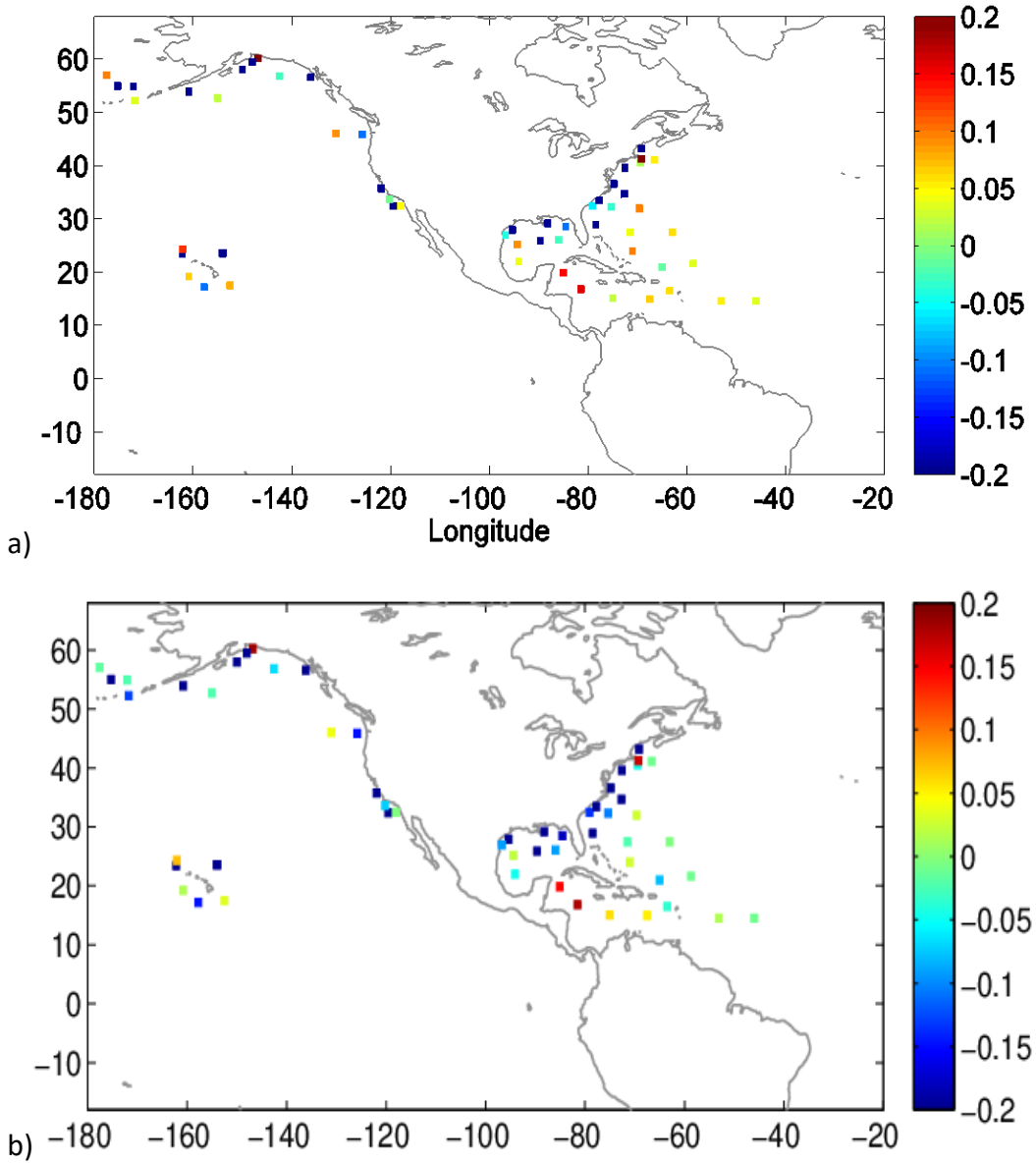

Figure 7. Normalized biases in $H_s$ (m) for the matchups, for case 2 (a), and case 3 (b).

**Tables**

| ASTD (°C) | $U_N$ (m/s) at 4m | $U_N$ (m/s) at 8m |
|:---:|:---:|:---:|
| -4 | 6.05 | 6.42 |
| -2 | 5.90 | 6.35 |
| 0 | 5.75 | 6.20 |
| 2 | 5.55 | 5.91 |
| 4 | 5.30 | 5.78 |

Table 1. Comparison of wind speed values in neutral (cero), stable (positive) and unstable (negative values of ASTD) conditions, at different heights within the SBL. $U_N$=neutral equivalent.